\documentclass[aps,prl,epsfig,floats,twocolumn,superscriptaddress,amssymb,amsmath,floatfix,showpacs]{revtex4}
\usepackage{natbib}
\usepackage{graphicx}
\usepackage{color}
\begin{document}
\title{An Opto-Electric Micropump for Saline Fluids}

\author{Reza Kiani Iranpour}
\affiliation{Department of Physics, University of Guilan, P.O. Box 41335-1914, Rasht, Iran}

\author{Seyyed Nader Rasuli}
\email{rasuli@ipm.ir}
\affiliation{Department of Physics, University of Guilan, P.O. Box 41335-1914, Rasht, Iran}

\date{\today}

\begin{abstract}
A novel method to pump fluid in lab on chip devices with velocities up to tens of micrometer per second is introduced. A focused laser beam locally heats up an electrolyte. A net charge tends to accumulate in the heat-absorbing area, due to unequal tendencies of positive and negative ions to move in the presence of the temperature gradient. An external electric field then exerts a net force on the accumulated charge and consequently on water. This causes flow of water, with velocities up to tens of micrometer per second, for a simple NaCl+water solution. The method lets us change direction and amount of fluid pumping, simply by replacing the focal area.
\end{abstract}


\maketitle

\paragraph{Introduction.} As science/technology goes forward the smallest available scale by which man can construct new or manipulate existing apparatus constantly decreases. Therefore, we should create smaller tools which function with enough accuracy, stability and speed. Thinking of a microfluidic setup\cite{Microfluidics}, we need apparatus which capture sub-micron particles suspending in fluid \cite{Light_Induced_Capturing}, or micro-pumps which pump femto/pico liters of fluid in the required direction with the needed speed\cite{Light_Induced_Pumping,Thermo-Viscous-Expansion}. The traditional picture of a pump, for example, is constructed out of many smaller segments. Inevitably, we have to miniaturize all its segments to obtain the same pump in micron scales; this seems a serious technical challenge. One solution is to change our approach to capturing particles or pumping micro/nano fluid \cite{Light_Induced_Capturing,Light_Induced_Pumping,Thermo-Viscous-Expansion}. Weinert {\em{et al.}} for example, have constructed a new generation of micro-pumps which make use of both a moving hot spot and the temperature dependence of fluid viscosity; they could pump fluid in the desired direction \cite{Thermo-Viscous-Expansion}.
\begin{figure}
\includegraphics[width=0.99\columnwidth]{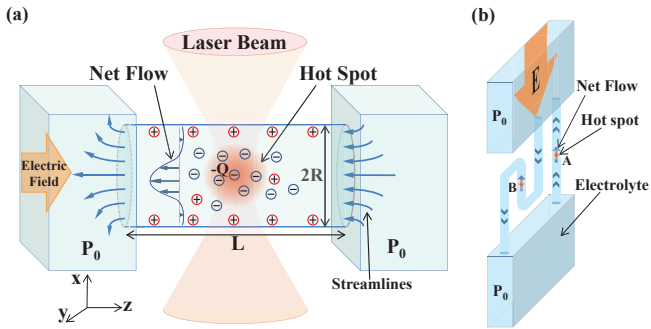}
\caption{(Color Online). {\bf{(a)}} A focused laser beam heats a small area in the middle of a tube connecting two equi-pressure containers. Unequal Soret coefficients of negative and positive ions cause accumulation of a net charge in the heated area and a neutralizing layer on tube's surface. An external electric field then brings fluid into motion. {\bf{(b)}} While the local direction of flow pumping depends on the external field, an "N" shaped tube lets us reverse its global direction. Here we can pump in either of directions by replacing the focal area from {\bf{A}} to {\bf{B}}.}
\label{Fig-suggested-pump}
\end{figure}

Here we similarly use a light induced hot spot together with an external electric field to propose another possible generation of micro-pumps. We assume an aqueous fluid with ions solved in it. Using a light induced heated area Fig. 1(left), we are able to create locally charged regions in it \cite{Alois-soft-matter,Reza-Nader}; an external electric field then exerts a net force on the ions---accumulated in the heat-absorbing area---and cause fluid to flow (see Fig. 1).

\paragraph{Charge density.} A focused laser beam heats a small area of an electrolyte and creates a local hot spot. First, to obtain a clearer picture, we approximate the light absorbing area to a small sphere of radius $a$; which {\em homogenously} absorbs energy with a total rate of $\dot{\Omega}_{abs}$. The heat conduction equation, in the steady sate, is: $-\tilde{\kappa}\nabla^{2}T(\vec{r})=\Theta(a-r)\times3\dot{\Omega}_{abs}/(4\pi a^3)$; where $\tilde{\kappa}$ is the thermal conductivity of solution, $T$ is the temperature, $r$ is distance from heated sphere's origin and $\Theta(x)$ the Heaviside step function. We read the temperature as:
\begin{eqnarray}
T(r)=T_{0}+ \delta T_{max}
\begin{cases}
 1 - r^2/3a^2 \text{,\qquad\qquad\quad} r\leq a\\
 2a/3r \text{,\qquad\qquad\qquad\quad} r>a.
\end{cases}
\label{Temperature-prof}
\end{eqnarray}
where $\delta T_{max}=3\dot{\Omega}_{abs}/8\pi a\tilde{\kappa}$. As we should expect, for $r>a$, Eq. (\ref{Temperature-prof}) presents a constant outward energy flow: $4\pi r^2 \times\tilde{\kappa}|\vec{\nabla}T(r)|=(8\pi a\tilde{\kappa}/3)\times\delta T_{max}=\dot{\Omega}_{abs}$. In other word, a $1/r$ tail for temperature decay is the inevitable consequence of energy conservation---as long as spherical symmetry of system is assumed.

Solute ions around this heated area feel a temperature gradient which causes their motion \cite{Ludwig}. The {\em{current density}} of the $i$th type of ions ({\em i.e. $\vec{J}_{i}$}) is read as \cite{de Groot}:
\begin{eqnarray}
\label{Diffusion-equation}
\vec{J}_{i}=- D_{i} \vec{\nabla}C_{i} - \mu_{i} q_{i} C_{i} \vec{\nabla}\phi - S_{T\!\mbox{-}i} D_{i} C_{i} \vec{\nabla}T.
\end{eqnarray}
Where $D_{i}$ and $C_{i}$ are the diffusion constant and number density of the $i$th type of ions; $\mu_{i}$ and $q_{i}$ are ion's mobility and charge; and $\phi$ is the electric potential induced by possible charge accumulation (note that there is no external electric field yet). $S_{T\!\mbox{-}i}$, known as {\em Soret coefficient}, is a parameter of dimension $T^{-1}$ which describes ions' motion in the presence of a temperature gradient \cite{de Groot}.
\begin{figure}
\includegraphics[width=0.99\columnwidth]{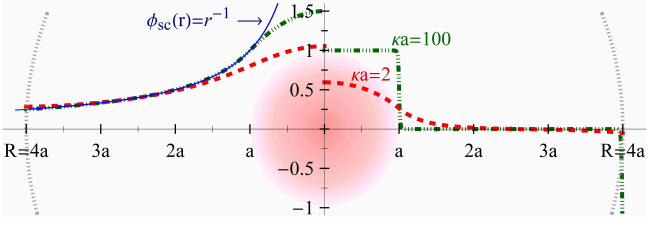}
\caption{(Color Online). Saline fluid is confined to a sphere of radius $R=4a$ (presented with dotted gray arcs). Fluid uniformly absorbs heat inside a sphere of radius $a$ (the circular pink area). {\bf{Left:}} the scaled potential ({\em i.e. }$\phi(r)\times\varepsilon_{w}Q_{tot}^{-1}$) plotted versus $r$. The dashed graph (red) plotted for $\kappa a=2$ while the dashed-dot-dot (green) graph represents $\kappa a=100$. Both curves converge to $\phi_{SC}=r^{-1}$, when $r-a$ becomes few times larger than $\lambda_{DH}=\kappa^{-1}$. {\bf{Right:}} the scaled corresponding charge density ({\em i.e. }$\rho_{ch}(r)\times 4\pi a^{3}/(3 Q_{tot})$). For $\kappa a=100$ the charge is almost homogenously accumulated in a sphere of radius $a$; besides, a neutralizing layer appears close to container's surface, at $R-\lambda_{DH} \leq r \leq R$. For $\kappa a=2$, the two oppositely charged regions still exist, but with much smother boundaries.}
\label{Fig-Charge-Potential}
\end{figure}

To obtain some feeling of charge accumulation, we follow relaxation of an unbounded water+NaCl solution immediately after laser beam is applied. In the absence of the beam, there is a homogenous solution $C_{+}=C_{-}=C_{0}$, thus no charge accumulation exists and $\phi=0$. Applying the laser beam increases the temperature around focal area and creates an inward temperature gradient (see Eq. {\ref{Temperature-prof}}). Experimental data show that for the NaCl solution in room temperature ({\em i.e.}$\,T_{0}=300^\circ$K) $S_{T\!\mbox{-}Na}\simeq0.7/T_{0}$ and $S_{T\!\mbox{-}Cl}\simeq0.1/T_{0}\,$\cite{Gaets}. Consequently both ions are pushed outward due to this inward temperature gradient (note $- S_{T\!\mbox{-}i} D_{i} C_{i} \vec{\nabla}T$ in $\vec{J}_{i}$). However, Na$^{+}$ ions have larger Soret coefficient; so while both ions' number densities are decreasing in the heated area, the number density of Sodium ions ({\em i.e.} $C_{+}$) is decreasing more rapidly than that of Chlorines' ({\em i.e.} $C_{-}$). Therefore a net negative charge begins to appear in the heated area.

In the steady state, both current densities should vanish ({\em i.e.} $J_{+}=J_{-}=0$) \cite{Reza-Nader}. Here, we consider increment in local temperature ({\em i.e.} $\delta T=T-T_{0}$) and its derivative as source of perturbation and follow a {\em{linear response}} approach. The steady state condition then simplifies to:
\begin{eqnarray}
\vec{J}_{i}= - D_{i0} \vec{\nabla}C_{i} - \mu_{i0} q_{i} C_{0} \vec{\nabla}\phi - S_{T\!\mbox{-}i} D_{i0 } C_{0} \vec{\nabla}\delta T=0.
\label{steady-condition}
\end{eqnarray}
The index $0$ indicates unperturbed values, when $\delta T=0$ everywhere. We use the Stokes-Einstein relation for diffusion $D_{i}=\mu_{i}K_{B}T$ and assume that far enough from the heated area: $\phi(\vec{r} \rightarrow \infty)=0$ and $\delta T(\vec{r}\rightarrow\infty)=0$ \cite{Reza-Nader}. Then Eq. (\ref{steady-condition}) yields: $C_{i}= C_{0}\{1-q_{i}\phi/(K_{B}T_{0})-S_{T\!\mbox{-}i}\delta T\}$. Now we can evaluate the charge density as:
\begin{equation}
\rho_{ch}\!=\!q(C_{+}-C_{-})\!=\!qC_{0}\{-\frac{2q\phi}{K_{B}T_{0}}-(S_{T+}-S_{T-})\delta T\},
\label{Rho-ch}
\end{equation}
where $q=q_{+}$ is the protonic charge. This result suggests that far from heat-absorbing area, where $\delta T \propto 1/r$ (see Eq. (\ref{Temperature-prof}) for $r>a$), the assumption of charge neutrality $\rho_{ch}=0$ gives an electric potential like:
\begin{eqnarray}
\phi=\frac{Q_{tot}}{\varepsilon_{w}r},
\text{\, with } Q_{tot}= -q\frac{a}{3 l_{B}}\times\delta T_{max}(S_{T+}-S_{T-}).
\label{Net-charge}
\end{eqnarray}
Where $\varepsilon_{w}$ is the electric permittivity of water, and $l_{B}=q^2/\varepsilon_{w} K_{B} T$ is the Bjerrum length \cite{colloid}. Eq. (\ref{Net-charge}) is signature of a net charge of $Q_{tot}$ accumulated around heat-absorbing area. A more careful treatment however requires us to solve the Poisson equation in whole space \cite{Jackson}:
\begin{eqnarray}
\vec{\nabla}.(\varepsilon_{w} \vec{\nabla}\phi)=-4\pi \rho_{ch}.
\label{Poisson}
\end{eqnarray}
The {\em{linear response approximation}} lets us neglect temperature dependence of $\varepsilon_{w}$ and take it as a constant. We merge Eq.(\ref{Rho-ch}) and Eq.(\ref{Poisson}) and obtain:
\begin{eqnarray}
(\nabla^2 - \kappa^2)\phi=\frac{4\pi q C_{0}}{\varepsilon_{w}} (S_{T+}-S_{T-})\delta T,
\label{thermoelectric-equation}
\end{eqnarray}
where $\kappa=\sqrt{(8\pi l_{B} C_{0})}=\lambda_{DH}^{-1}$ is the inverse of Debye screening length \cite{colloid}. This result, also found by Majee and W\"{u}rger \cite{Alois-soft-matter}, describes electric potential in the presence of a varying temperature field. We solve it for a mixture of water and a 1:1 salt, which is confined in a rigid sphere of radius $R$ and absorbs light in a smaller sphere of radius $a$ (Fig. \ref{Fig-Charge-Potential}). The global charge neutrality yields $\partial_{r}\phi|_{r=R}=0$; and the absence of any singular charge in origin imposes: $\lim_{r \rightarrow 0} r^{2} \partial_{r}\phi=0$ \cite{Reza-Nader}. We solve Eq. (\ref{thermoelectric-equation}) with these boundary conditions and get $\phi(r)$ as:
\begin{eqnarray}
\!\frac{Q_{tot}}
{\varepsilon_{w}r}\!\times\!
\begin{cases}
r(6\kappa^{-2}+r^2\!-3a^2)/(2a^3)-{\tilde{\gamma}}\sinh(\kappa r)\text{,}\,\; r\leq a\!\!\!\\ 1\!+\!{\tilde{\upsilon}}\exp(-\kappa r)\!+\!{\tilde{\omega}}\exp(-\kappa\{R-r\}) \text{,  } a\!\leq\! r\!\leq\! R,\!\!\!
\label{Eq-potential}
\end{cases}
\end{eqnarray}
where ${\tilde{\gamma}}=3e^{-\kappa a}(1+\kappa a)/(\kappa a)^3$, ${\tilde{\omega}}=a/(1-\kappa R)$, and ${\tilde{\upsilon}}=3\{\sinh(\kappa a)-\kappa a \cosh(\kappa a)\}/(\kappa a)^3$ (here we have neglected corrections like $\exp(-\kappa R)$, in ${\tilde{\gamma}}$, ${\tilde{\omega}}$, and ${\tilde{\upsilon}}$).
Figure \ref{Fig-Charge-Potential}(left) shows two (scaled) electric potential as function of $r$. Both curves, corresponding to $\lambda_{DH}=a/100$ and $\lambda_{DH}=a/2$, converge to the leading curve of $\phi_{sc}(r)=1/r$ when $r\!-\!a\gtrsim\lambda_{DH}$.

The right side of Figure \ref{Fig-Charge-Potential} shows the (scaled) charge density using electric potential in Eq. (8). Both curves show an accumulation of charge around heat-absorbing area. Particularly for $\kappa a =100$ the charge is homogenously accumulated with in the heat-absorbing area ({\em{i.e.}} $\rho(r<a)\simeq1$), conceiving a {\em{homogenously}} charged spherical fluid (CSF) of radius $a$ \cite{Reza-Nader}. Besides, global neutrality forms a neutralizing layer with few $\lambda_{DH}$ thickness on the inner surface of container ({\em{i.e.}} $R-\lambda_{DH} \leq r \leq R$).

In our suggested pump, however, the saline fluid is not confined to a sphere but to a glassy tube with length $L$ and radius $R$ (Fig. 1). This breaks the spherical symmetry and shrinks CSF into a prolate with larger axial ({\em i.e.} the z direction) radius. Fortunately, for $\lambda_{DH} \ll R-a$\cite{loenly-focus}, the picture of a {\em spherical} homogenously charged fluid is still applicable ({\em e.g.} for $\kappa a=100$ and $R=2a$ the prolate's elipticity is $(r_{max}-r_{min})/r_{min} \lesssim 10^{-8}$) \cite{Reza-Nader}. The neutralizing layer also, forms beside inner surface of the tube, with a charge density:$\,\rho_{ch}(\rho, z)=\sigma_{w}(z)\times\kappa\exp(-\kappa\{R-\rho\})$\cite{tube-suface-charge}. $\sigma_{w}(z)$ decays fast for $|z|>R$; this means that almost whole neutralizing charge is gathered in a cylinder of length $L=2R$ \cite{Reza-Nader}. Consequently, the charge densities in a tube with $L \geq 2R$, are almost the same as charge densities in an infinite tube.
\begin{figure}
\includegraphics[width=0.99\columnwidth]{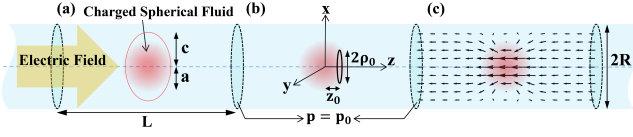}
\caption{(Color Online). A chain of tubes in an external field of $E_{0}\hat{z}$. The fluid pressure has its bulk value of $P_{0}$ at all interfaces between successive tubes (shown by black dotted ellipses). ({\bf{a}}) A tube with charged spherical fluid inside it. The full line (purple) ellipse conceives possible extension of focal area to a spheroid (see Eq.(\ref{eq-Elliptic-Charge})). ({\bf{b}}) Depiction of used coordinate system and Green's function (see Eq.(\ref{eq-Green})). ({\bf{c}}) Full fluid velocity field, obtained using Green's function\cite{Reza-Nader}.}
\label{Fig-repeat-spot}
\end{figure}

\paragraph{Fluid flow.}
So there are two charged regions, a homogenously charged spherical fluid and a neutralizing layer, both inside the tube which connects two containers with equal pressure of $P_{0}$ (Fig. \ref{Fig-suggested-pump}). An axial electric field then exerts a body force of $\rho_{ch}(\vec{r})E_{0}\hat{z}$ on these charged regions and, consequently, on fluid. However ions in the neutralizing layer are much closer to tube's surface; they feel a far larger hydrodynamic friction and loss game to ions inside CSF. We address fluid steady motion with Stokes equation, providing {\em{low Reynolds number}} assumption \cite{Happel-Brenner}:
\begin{eqnarray}
0 = -\vec{\nabla}P(\rho, z)+\eta \nabla^{2}\vec{V}(\rho, z) + \rho_{ch}(\rho, z)E_{0}\hat{z}.
\label{Stokes-equation}
\end{eqnarray}
For two ends of the tube, the boundary conditions are $P(\rho,L/2)=P(\rho,-L/2)=P_{0}$. On the tube's lateral surfaces, fluid has zero normal component:$V_{\rho}(R,z)=0$; and a non-zero tangential component which is determined by the slip length of $b$ as: $V_{z}(R,z) = b \times \partial_{\rho}V_{z}(\rho,z)|_{\rho\rightarrow R^{-}}$.

For a long enough tube, we may neglect the effect of fluid's complicated motion around {\em{tube-container junction}} ({\em{e.g.}} complicated distancing/closing stream lines at two ends of the tube in Fig.(\ref{Fig-suggested-pump})). This lets us assume a chain of tubes, with mentioned charged regions inside all of them (see Fig.\ref{Fig-repeat-spot}). An external electric field of $E_{0}\hat{z}$ is applied to this system, and the condition of fixed pressure at the interfaces between successive tubes ({\em{i.e.}}$\,P(\rho,\,z=n\times L+L/2)=P_{0}$) is considered.
With {\em{tube-container junction}} neglected, the fluid velocity field in each of the repeating tubes would be the same as in the single tube of Figure (\ref{Fig-suggested-pump}). We use this picture of repeating tubes to expand fluid velocity/pressure in Fourier series:
$\vec{V}(\rho, z)=\sum_{n = -\infty}^{\infty}\vec{U}(\rho, n)\times\exp(2\pi i n z/L)$ \cite{Happel-Brenner}; this lets us write/solve proper Green's equation with cylindrical symmetry (Fig.\ref{Fig-repeat-spot}{\bf b}):
\begin{equation}
0=-\vec{\nabla}P_{\text{\tiny{G}}}+ \eta \nabla^{2} \vec{V}_{\text{\tiny{G}}}+ F_{0}\hat{z} \frac{\delta(\rho-\rho_0)}{2 \pi \rho_0}\!\sum_{n = -\infty}^{\infty}\!\delta(z-z_{0}-nL),
\label{eq-Green}
\end{equation}
where, in all tubes, an axial force of $F_{0}\hat{z}$ is uniformly imposed to a circle of radius $\rho_{0}$ which sits $z_{0}$ ahead of tube's center. We use Fourier series to calculate $V_{\text{\tiny{G}}}$ and $P_{\text{\tiny{G}}}$ \cite{Reza-Nader}; using $V_{\text{\tiny{G}}}$ and numerical integration over force distribution in the tube, gives us the fluid velocity field (Fig.\ref{Fig-repeat-spot}{\bf{c}}).

However our objective is to calculate the net current induced in the tube: $I(z)=\int_{0}^{R} 2\pi \rho \, d\rho V_{z}(\rho, z)$. For an {\em incompressible fluid}, $I(z)$ is independent of $z$; therefore the $z$ component of {\em the only $z$ independent term} in fourier expansion of $\vec{V}(\rho, z)$ ({\em i.e.}$\;U_{z}(\rho,n\!=\!0)$) takes part in it \cite{z-independence}. Similarly, to calculate $I_{G}(\rho_0)$, the current induced by force distribution in Eq.(\ref{eq-Green}), we only need $U_{\text{\tiny{G}}z}(\rho,n=0)$ which is the $z$ component of $n=0$ fourier term of $\vec{V}_{G}$:
\begin{equation}
0=\eta \frac{1}{\rho} \partial_{\rho} \{\rho\,\partial_{\rho}\,U_{\text{\tiny{G}}z}(\rho,0)\}+{F}_{0} \frac{\delta(\rho-\rho_0)}{2 \pi \rho_0}\times\frac{1}{L}.
\label{Eq-govenr-Uz0}
\end{equation}
Here $-\vec{\nabla}P_{\text{\tiny{G}}}$ hasn't any zeroth order fourier term, due to the applied periodicity on pressure. We solve Eq.(\ref{Eq-govenr-Uz0}) for $U_{\text{\tiny{G}}z}(\rho,0)$ and calculate $I_{G}(\rho_0)$ as:
\begin{equation}
I_{\text{\tiny{G}}}(\rho_0)\!=\!\int_{0}^{R}\!2\pi \rho \, d\rho\, U_{\text{\tiny{G}}z}(\rho,0)=\frac{F_{0}}{4 \eta L} \{(R^{2}-\rho_{0}^{2}) + 2 b R\}.
\label{eq_Green_Current}
\end{equation}
This current is maximum when $F_{0}\hat{z}$ is applied to a point on tube's axis ({\em i.e.}\,$\rho_{0}\!=\!0$). It monotonically falls to its minimum in $\rho_{0}=R$, where force is applied just adjacent to tube's surface. The minimum is portion of finite slip length: $I_{slip}=F_{0}\,b\times(R/2\eta L)$ and independent of $\rho_{0}$. In other words, $I_{slip}$ only depends on total force and not on its distribution.

We integrate $I_{\text{\tiny{G}}}$ over whole force distributions inside tube to obtain total current induced in the opto-electric pump. For a long enough tube ($L \geq 2R$), the charged spherical fluid and neutralizing layer have {\em{equal}} and opposite charges. They feel equal and opposite forces; thus their induced $I_{slip}$s would be {\em{equal}} and opposite, and cancel each other. This eliminates the slip portion in the net current. However, the rest of the induced current (obtained by integration over $F_{0}\times (R^{2}-\rho_{0}^{2})/(4\eta L)$) is different: the neutralizing layer has a negligible portion (as $R-\lambda_{DH} \leq \rho_{0} \leq R$) while the CSF has a remarkable portion, which is almost total induced current:
\begin{equation}
I_{tot}=\pi R^{2}\times \frac{Q_{tot} E_{0}}{4\pi \eta L} \{1-\frac{2}{5}\frac{a^2}{R^{2}}\}.
\label{eq-I-tot}
\end{equation}
Then, the {\em{mean fluid velocity}} $\bar{V}=I_{tot}/\pi R^{2}$ is:
\begin{equation}
\bar{V}=\frac{Q_{tot} E_{0}}{4\pi \eta L} \times \{1-\frac{2}{5}\frac{a^2}{R^{2}}\}.
\label{eq-V-drift}
\end{equation}
Here the first part of Eq.(\ref{eq-V-drift}) resembles the drift velocity of an assumed suspending particle with charge of $Q_{tot}$ and radius of $2L/3$, which feels an electric field of $E_{0}$. The 2nd part ({\em i.e.}$1-2a^2/5R^{2}$) is a {\em{shape correction}} due to the dispersion of $Q_{tot}$ in a spherical fluid of radius $a$ instead of accumulating in a point on tube's axis ({\em i.e.}$\,a=0$).

In order to check the reliability of our results, which are based on the spherical focal area approximation, we approximate the focal area to a spheroid which has a larger radius of $c$ parallel to light propagation axis, and smaller radiuses of $a$ in two other axes (see Fig.3{\bf{a}}). The spheroid {\em{homogenously}} absorbs light; with $\lambda_{DH}\ll R-c$ \cite{loenly-focus}, a homogenously charged {\em{spheroidal}} fluid appears \cite{Reza-Nader}. This corrects $Q_{tot}$ (in Eq.(\ref{Net-charge})) by a multiplicand of \cite{Reza-Nader,Spheroidal_correction}:
\begin{equation}
f(c/a)=\frac{\sqrt{c^2/a^2-1}}{\sinh^{-1}(\sqrt{c^2/a^2-1})}.
\label{eq-Elliptic-Charge}
\end{equation}
The charged spheroidal fluid, also, corrects the {\em{shape correction}} in Eq.(\ref{eq-V-drift}) to $\{1-(a^2+c^2)/5R^{2}\}$ \cite{Reza-Nader}. These two corrections increase $\bar{V}$, but do not change its order of magnitude\cite{Reza-Nader}.

\paragraph{Discussion.} Now, we imagine a tube of length $L=8\mu$m and radius $R=2\mu$m (see Fig.(3)), which is filled with saline fluid ($\kappa \simeq 1/nm$) and feels an external electric field of $E_{0}=450V/\text{cm}$. Light beam heats up a sphere of radius $a=1\mu$m in its middle so that temperature in tube's center rises by $\delta T_{max}=15^\circ$C. This causes a total charge of $Q_{tot}\approx-14\times q$, to homogenously accumulate in the hot spot. Using Eq.(\ref{eq-V-drift}), the mean fluid velocity would be obtained as $\bar{V}=1\mu\text{m}/\text{s}$, opposite to the applied electric field. If a spheroidal focal area is considered with $a \leq c \leq R$, the estimated mean fluid velocity would rise by a factor between $1$ and $1.1$\cite{Reza-Nader}.

It is tempting to look for any mean to increase the fluid velocity of a pump which functions at room temperature ($T_{0}\approx 293^\circ$C). The geometrical parameters do not seem to be much flexible as we have made certain assumptions like a long enough tube ({\em i.e} $L > 2R$) and highly saline regime ({\em i.e.} $\kappa a \gg 1)$. The only flexible variables seem to be $\delta T_{max}$ and $E_{0}$. In general we have to avoid fluid vaporization; this bounds $\delta T_{max}$ by something like $60^\circ$C. For a fluid which contains proteins and even living materials, $\delta T_{max}$ should be much lower, like $15^\circ$C. A similar story holds for the electric field, as its value for a simple NaCl+Water solution can go up to $5\text{kV}/\text{cm}$, while with living materials it has to be few hundreds of volt per cm\cite{Cell_Joule_Heatting}. These yield velocities up to $50\mu\text{m}/\text{s}$ for simple water+salt solution, and a velocity about $1\mu\text{m}/\text{s}$ for living matter. However, due to the short passage time of a living cell through focal area, we hope that even $\delta T_{max}$s higher than $15^\circ$C could be tolerated by living matter.

It seems evident that the pumping speed would reduce if whole focal area moves toward tube's lateral surface \cite{Reza-Nader}. However, we may also reverse the direction of fluid flow, using a simple trick of geometry. Fig.1{\bf{b}} shows how replacing focal area, from {\bf{A}} to {\bf{B}}, reverses the upward global direction of fluid transport to downward. However this nice feature may be jeopardized by possible long {\em relaxation time} that pump needs to become operative/inoperative, after replacement of the focal area. There are three time scales corresponding to three stages pump needs to become operative: the heat diffusion time ({\em i.e. }$\tau_{heat}\simeq R^2/D_{heat}$) that scales formation of temperature field, the ions' diffusion time ({\em i.e. }$\tau_{ions}\simeq R^2/D$) which scales formation of {\em CSF}/{\em neutralizing layer}, and finally the hydrodynamics relaxation time ({\em i.e. }$\tau_{hydro}\simeq \rho L^2/\eta$) which measures fluid's evolution toward steady motion \cite{Reza-Nader}. Considering the dimensions of the pump we gave as example, these time scales would be: $\tau_{heat}\simeq 0.02\text{ms}$, $\tau_{ions}\simeq 3\text{ms}$, and $\tau_{hydro}\simeq 0.06\text{ms}$; all appears to be short enough for practical fluid pumping in lab on chip setups.

In summary, we suggest a new method for pumping saline fluid, which works with combination of an external electric field and a focused light beam, and pumps fluid with velocities up to $50\mu\text{m}/\text{s}$. For a fixed external field, the fluid motion could be simply tuned/reversed by changing the position of the focal area.

We are grateful to M.R.Ejtehadi, S.Rashid-Shomali, R.Golestanian, E.Branguier and S.N.S.Reihani for careful and stimulating comments. S.N.R. acknowledges financial support provided by Iran National Foundation of Elites, through Young Professors Grant program.



\begin{thebibliography}{99}
\bibitem{tube-suface-charge}
The relaiton: $\rho_{ch}(\rho, z)=\sigma_{w}(z)\times\kappa\exp(-\kappa\{R-\rho\})$ is obtained for the $\lambda_{DH} \ll R-a$ limit. Here we may consider $\sigma_{w}(z)$ as surface charge of the thin neutralizing layer.
\bibitem{z-independence}
Using fluid incompressibily $\vec{\nabla}.\vec{V}=0$, we can directly show that only $U_{z}(\rho, 0)$ has non-zero portion in total current.
\bibitem{loenly-focus}
This means that focal ara is far enough from cylinder's lateral surface.
\bibitem{Spheroidal_correction}
For $1\leq c/a \leq 3$, $f(c/a)$
varies between $1$ and $1.6$ and is well approximated to $(2/3)+(c/3a)$.
\end{thebibliography}
\end{document}